\documentclass[extra]{gji_arxiv}
\usepackage{timet, color}
\usepackage{amsmath, amssymb, bm, cuted, mathtools}
\usepackage{layouts}
\usepackage{graphicx}
\usepackage{wrapfig}
\usepackage[hidelinks]{hyperref}
\renewcommand{\Re}{\operatorname{Re}}
\renewcommand{\Im}{\operatorname{Im}}
\DeclareMathOperator{\arctantwo}{arctan2}

\title[6C fingerprinting]
  {Efficient wave type fingerprinting and filtering by six-component polarization analysis}
\author[D. Sollberger  {\normalfont et al.}]
  {David Sollberger$^1$, Nicholas Bradley$^{1,2}$, Pascal Edme$^1$, Johan O. A. Robertsson$^1$ \\ \\ 
  $^1$ Institute of Geophysics, ETH Zurich, Zurich, Switzerland \\
  $^2$ Now at: Centre for Geophysical Forecasting, NTNU, Trondheim, Norway
  }
\date{November 2022}
\pagerange{\pageref{firstpage}--\pageref{lastpage}}
\pubyear{2022}

\begin{document}

\label{firstpage}

\maketitle

\begin{summary}
We present a technique to automatically classify the wave type of seismic phases that are recorded on a single six-component recording station (measuring both three components of translational and rotational ground motion) at the earth's surface. We make use of the fact that each wave type leaves a unique 'fingerprint' in the six-component motion of the sensor. This fingerprint can be extracted by performing an eigenanalysis of the data covariance matrix, similar to conventional three-component polarization analysis. To assign a wave type to the fingerprint extracted from the data, we compare it to analytically derived six-component polarization models that are valid for pure-state plane wave arrivals. For efficient classification, we make use of the supervised machine learning method of support vector machines that is trained using data-independent, analytically-derived six-component polarization models. This enables the rapid classification of seismic phases in a fully automated fashion, even for large data volumes, such as encountered in land-seismic exploration or ambient noise seismology. Once the wave-type is known, additional wave parameters (velocity, directionality, and ellipticity) can be directly extracted from the six-component polarization states without the need to resort to expensive optimization algorithms.

We illustrate the benefits of our approach on various real and synthetic data examples for applications such as automated phase picking, aliased ground-roll suppression in land-seismic exploration, and the rapid close-to real time extraction of surface wave dispersion curves from single-station recordings of ambient noise. Additionally, we argue that an initial step of wave type classification is necessary in order to successfully apply the common technique of extracting phase velocities from combined measurements of rotational and translational motion.
\end{summary}


\section{Introduction}
Three-component polarization analysis and filtering is widely used in seismology and seismic exploration to characterise seismic wave motion in terms of its directionality and ellipticity, and to  enhance the signal-to-noise ratio of specific seismic phases \citep[e.g.,][]{Flinn1965, Vidale1986, Montalbetti1970, Christoffersson1988, Greenhalgh2018}. It has recently been shown that some of the ambiguity that is inherent to three-component polarization analysis can be overcome when data from rotational seismometers are additionally included in the analysis, leading to so-called six-component polarization analysis schemes. For example, rotational data help to eliminate the $180^{\circ}$ ambiguity that is inherent to three-component direction finding problems and enables the unambiguous estimation of the sense of rotation (ellipticity angle) of Rayleigh waves in the single station case \citep{Marano2013, Sollberger2018, Sollberger2020}. 

Another key advantage of incorporating rotational data into polarization analysis is that it enables the estimation of the seismic wave type of a recorded polarized wave. While at a single point of observation, pure translational motions are not indicative of the wave type (e.g., the polarization of both P- and S-waves is rectilinear in the three translational components), rotational motion recordings are a direct measurement of the wavefield’s curl and are thus exclusive to S-waves and surface waves in isotropic media. Hence, rotational motion measurements can highly facilitate the identification and isolation of different wave modes in seismic data. The concept of computing the wavefield divergence and curl from small-aperture arrays has been widely exploited to separate P- from S-waves \citep{Robertsson1999, Robertsson2002, Woelz2005, Sollberger2016a, VanRenterghem2018} and to locally identify and suppress surface waves in land seismic exploration data \citep{Edme2013, Edme2013a, Barak2014, Muyzert2019, Allouche2020}. The wave type selectivity provided by rotation data has also been used to analyse Love waves in the secondary microseism \citep{Hadziioannou2012, Tanimoto2015, Tanimoto2016} since Love (or SH-waves) are the only wave types that generate rotational motions around the vertical axis. \citet{Yuan2020} developed a polarization analysis scheme for back-azimuth estimation that only makes use of the horizontal components of a rotational seismometer which are purely sensitive to SV- and Rayleigh waves, thereby enabling a stable estimation of the back-azimuth that does not suffer from the interference of other wave types \citep{Yuan2021}. \citet{Sollberger2018} and \cite{Sollberger2020} introduced a method to infer on the wave type of recorded six-component ground-motion by fitting analytically derived wave models to the data. The method has been recently used to characterize the composition of the seismic wavefield at Stromboli volcano \citep{Wassermann2022}. Since the method relies on an expensive grid search for the wave parameters, it is not feasible to apply it to large data volumes.

In the present paper, we introduce an efficient way to identify the wave types of seismic phases that are recorded on a single six-component recording station (measuring both translation and rotation) at the earth's surface. To do so, we extract time- and frequency-dependent 6C polarization states of seismic wave arrivals by performing an eigenanalysis of the complex covariance matrix of the six component data. We then apply the machine learning technique of support vector machines to rapidly classify the wave type of an extracted polarization state represented by the principal eigenvector of the covariance matrix. The machine learning model is trained using purely analytical polarization models that are valid for isotropic media, resulting in a fully data-independent training. 

After the classification of the wave type of a recorded polarization state, dedicated filters can be defined that separate the data into different wave modes. This can highly facilitate the interpretation of single-station seismograms and enables the automated picking of seismic phases (P-, S-, and surface waves). Additionally, once the wave type is known, wave parameters like the local propagation velocity and the propagation direction can be estimated directly from the extracted polarization states without the need to employ costly grid-search algorithms and without the ambiguity inherent to three-component polarization analysis. 

\section{Theory}
We consider a simple seismic wave model $\mathbf{u}(\mathbf{r},t):\mathbb{R}^4 \to \mathbb{R}^6$ for a pure-state seismic wave of amplitude $A$ and angular frequency $\omega$ travelling in the direction of the wave vector $\mathbf{k}=[k_x, k_y, k_z]^T$. The magnitude of the wave vector is the wave number. The wave is described at position $\mathbf{r}=[x,y,z]^T$ and at time $t$ with the following plane-harmonic solution of the wave equation:

\begin{equation} 
 \label{eq:1-1}
{\mathbf{u}(\mathbf{r},t)}=\Re\left(A\mathbf{h}(\bm{\theta}, \omega)e^{j\left(\mathbf{k} \cdot \mathbf{r} - \omega t\right)}\right).
\end{equation}
Here, ${\mathbf{u}(\mathbf{r},t)}=(\dot{u}_x, \dot{u}_y, \dot{u}_z, {\omega}_x, {\omega}_y, {\omega}_z)^T(\mathbf{r},t)$, where the first three components of the six-component vector field describe the translational motions (in ground velocity) and the last three the rotational motions (in radians)\footnote{In practice, the actual measurements in the field can also be in acceleration (for the translational components) and rotation rate (for the rotational components). The processing techniques described in this paper can be applied to such data without loss of validity.}. $\Re(.)$ denotes the real part and $j$ is the imaginary unit with characteristic identity $j=\sqrt{-1}$.  The rotational motions are related to the curl of translational motions  as $(\omega_x,\omega_y,\omega_z)^T = \frac{1}{2} \nabla \times (u_x,u_y,u_z)^T$ \citep[e.g.,][]{Cochard_etal_2006}. In Eq.~\ref{eq:1-1}, the complex-valued vector quantity $\mathbf{h}(\bm{\theta}, \omega) \in \mathbb{C}^{6}$ describes the relative amplitudes and phase-shifts of the six-component motion and is in the following referred to as the 6-C polarization vector. It depends on a set of parameters $\bm{\theta}$, such as the mode of vibration (wave type), the propagation direction, and the local wave velocity and can be frequency-dependent (e.g., for dispersive surface waves). 

\begin{figure}
    \centering
        \includegraphics[width=0.6\columnwidth]{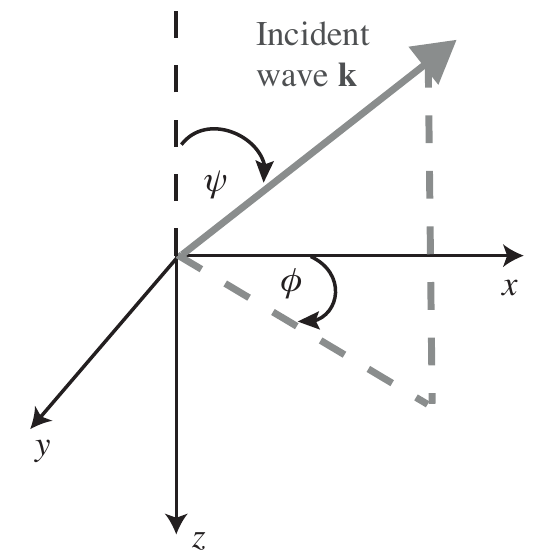}
    \caption{Right-handed coordinate frame considered in this paper. The propagation direction of a wave propagating in the direction of the wave vector $\mathbf{k}$ can be described with the inclination angle $\psi$ and the azimuth $\phi$.} 
    \label{fig:1-1}
\end{figure}

At the free-surface and for isotropic elastic media, the 6-C polarization vectors can be analytically derived for different wave types \citep[see][for a full derivation]{Sollberger2018}. In the following, we will consider a right-handed coordinate frame with a down-ward pointing z-axis as depicted in Fig.~\ref{fig:1-1}, consistent with the convention used by most commercial seismometers. The pure-state 6-C polarization vectors then take on the following expressions for P-, SV-, SH-, Rayleigh- and Love-waves, respectively \citep{Sollberger2018}:

\begin{equation}
    \begin{split}
         \label{eq:1-2}
            &{\bf{h}}^P(\alpha, \beta, \psi, \phi) = \\
            &\left( {\begin{array}{*{10}{c}}
            { -\cos \phi \left(\sin \psi + \frac{{{A_{PP}}}}{{{A_P}}}\sin \psi + \frac{{{A_{PS}}}}{{{A_P}}}\sqrt{1-\left(\frac{\beta}{\alpha}\sin{\psi}\right)^2}\right)}\\
            {  -\sin \phi \left( \sin \psi + \frac{{{A_{PP}}}}{{{A_P}}}\sin \psi + \frac{{{A_{PS}}}}{{{A_P}}}\sqrt{1-\left(\frac{\beta}{\alpha}\sin{\psi}\right)^2}\right)}\\
            { \cos \psi - \frac{{{A_{PP}}}}{{{A_P}}}\cos \psi + \frac{{{A_{PS}}}}{{{A_P}}}\left(\frac{\beta}{\alpha}\sin{\psi}\right)}\\
            {{{(2\beta )}^{ - 1}}\frac{{{A_{PS}}}}{{{A_P}}}\sin \phi }\\
            { - {{(2\beta )}^{ - 1}}\frac{{{A_{PS}}}}{{{A_P}}}\cos \phi }\\
            {0}
            \end{array}} \right),
            \end{split}
\end{equation} 

\begin{equation} 
    \begin{split}
        \label{eq:1-3}
    &{\bf{h}}^{SV}(\alpha, \beta, \psi, \phi) = \\
    &\left( {\begin{array}{*{10}{c}}
    {\cos \phi \left(\cos \psi - \frac{{{A_{SS}}}}{{{A_S}}}\cos \psi - \frac{{{A_{SP}}}}{{{A_S}}}\frac{\alpha}{\beta}\sin \psi\right)}\\
    {\sin \phi \left(\cos \psi - \frac{{{A_{SS}}}}{{{A_S}}}\cos \psi - \frac{{{A_{SP}}}}{{{A_S}}}\frac{\alpha}{\beta}\sin \psi\right)}\\
    {\sin \psi + \frac{{{A_{SS}}}}{{{A_S}}}\sin \psi - \frac{{{A_{SP}}}}{{{A_S}}}\sqrt{1-\left(\frac{\alpha}{\beta}\sin \psi\right)^2}}\\
    {{{(2\beta )}^{ - 1}}(1 + \frac{{{A_{SS}}}}{{{A_S}}})\sin \phi }\\
    { - {{(2\beta )}^{ - 1}}(1 + \frac{{{A_{SS}}}}{{{A_S}}})\cos \phi }\\
    0
    \end{array}} \right),
    \end{split}
\end{equation} 

\begin{equation} 
             \label{eq:1-4}
                {\bf{h}}^{SH}(\beta, \psi, \phi) = \left( {\begin{array}{*{20}{c}}
                {2\sin \phi }\\
                { - 2\cos \phi }\\
                0\\
                0\\
                0\\
                {-{\beta ^{ - 1}}\sin {\psi}}
                \end{array}} \right),
\end{equation}
\begin{equation}
            \label{eq:1-5}
                {{\bf{h}}^R}(c_R, \xi, \phi) = \left( {\begin{array}{*{20}{c}}
                {-j\sin \xi \cos \phi }\\
                {-j\sin \xi \sin \phi }\\
                { \cos \xi }\\
                {{c_R ^{ - 1}}\cos \xi \sin \phi }\\
                { - {c_R ^{ - 1}}\cos \xi \cos \phi }\\
                0
                \end{array}} \right),
\end{equation}

\begin{equation}
            \label{eq:1-6}
                {{\bf{h}}^L}(c_L, \phi) = \left( {\begin{array}{*{20}{c}}
                {2\sin \phi }\\
                { -2\cos \phi }\\
                0\\
                0\\
                0\\
                {-{c_L ^{ - 1}}}
                \end{array}} \right).
\end{equation}

In Eqs~(\ref{eq:1-2})--(\ref{eq:1-6}), $\psi$ denotes the inclination angle of the incident wave and $\phi$ the propagation azimuth (see Fig.~\ref{fig:1-1}). $\alpha$ and $\beta$ are the local P- and S-wave velocities at the recording station. $c_R$ and $c_L$ are the Rayleigh-wave and Love-wave phase velocity, respectively, and $\xi$ is the Rayleigh-wave ellipticity angle, determining the eccentricity and the sense of rotation of the Rayleigh-wave particle motion. 

For an incoming P-wave at a free surface, the polarization vector is a superposition of the three polarization vectors of the up-going, incident P-wave with amplitude $A_P$, the reflected down-going P-wave with amplitude $A_{PP}$, and the mode-converted down-going SV-wave with amplitude $A_{PS}$. Explicit expressions for the amplitudes of the superimposed waves as a function of the incident angle and the local P- and S-wave velocities can be found in Appendix~\ref{polarisationmodel}. Note that the non-zero rotation components in the P-wave polarization vector purely arise from the down-going mode-converted SV-wave.

Similarly, the polarization vector for an SV-wave is a superposition of the three polarization vectors of the up-going, incident SV-wave with amplitude $A_S$, the reflected down-going SV-wave with amplitude $A_{SS}$, and the mode-converted down-going P-wave with amplitude $A_{SP}$, where explicit expressions for the amplitudes can again be found in Appendix~\ref{polarisationmodel}. 

For overcritical SV-waves and Rayleigh waves, the imaginary part of the 6C polarization vectors is non-zero ($\Im({\mathbf{h}})\neq0$), which means that the motion in the six components is out of phase, describing an ellipse in 6D space. For all other wave types, the 6C polarization is rectilinear (the motion of all six components is in phase).  

Note that each polarization vector in Eqs~(\ref{eq:1-2})--(\ref{eq:1-6}) provides a template for a unique 'fingerprint' for each wave type with the exception of SH- and Love-waves, which have the same fingerprint (a Love-wave is nothing else but a horizontally propagating SH-wave). In the following, we attempt to extract this fingerprint from a 6C recording using polarization analysis and use it to automatically classify the wave mode of a recorded seismic phase.

\subsection{Measurement model}

A local six-component measurement $\mathbf{d}(\mathbf{r}_0, t)$ of a seismic wavefield  at position $\mathbf{r}_0$ is a superposition of an arbitrary number $n$ of pure-state waves and is typically corrupted by additive noise $\mathbf{n}(t)$, so that

\begin{equation}
    \label{eq:1-7}
    \mathbf{d}(\mathbf{r}_0, t) = \sum_{i=1}^{n}{\mathbf{u}_i(\mathbf{r}_0, t, \bm{\theta}_i)} + \mathbf{n}(t),
\end{equation}
where the character of the noise can be different on each of the six channels. Given some six-component data $\mathbf{d}(\mathbf{r}_0, t)$, we now want to infer on the polarization states of the recorded waves. 

Initially, the translational and rotational recordings have different units: ground velocity in m/s and rotational displacement in radians, respectively. For polarization analysis, it is convenient to convert the translational recordings to a dimensionless 'pseudo' rotation by scaling them by a slowness $p$, so that the resulting amplitudes of all six data components are comparable. This leads to the normalized data $\tilde{\mathbf{d}}(\mathbf{r}_0, t) = [p \dot{u}_x, p \dot{u}_y, p \dot{u}_z, \omega_x, \omega_y, \omega_z]^T(\mathbf{r}_0, t)$\footnote{Again, if the actual field measurements are in acceleration and rotation rate, the scaling slowness is simply applied to the acceleration components.}. The scaling slowness $p$ can hereby be chosen arbitrarily but it should ensure that the translational and rotational recordings have comparable amplitudes, warranting that the polarization analysis is numerically stable. In the following applications, we automatically select a scaling slowness based on the Euclidian norm of the translational and the rotational recordings as

\begin{equation}
    \label{eq:1-8}
    p = \frac{\displaystyle \int_{\mathbb{R}}{\left \lVert \left( \omega_x(t), \omega_y(t), \omega_z(t)\right)^T \right \rVert_2} dt}{\displaystyle \int_{\mathbb{R}}{\left \lVert \left(\dot{u}_x(t), \dot{u}_y(t), \dot{u}_z(t) \right)^T \right \rVert_2} dt}.
\end{equation}

\subsection{Polarization analysis in the time-frequency domain}
Since all parameters of the 6C polarization vectors listed in Eqs~(\ref{eq:1-2})--(\ref{eq:1-6}) can be frequency-dependent and the polarization states vary with time (due to the arrival of different waves), it is advantageous to estimate the 6C polarization states from a time-frequency representation of the recorded signal \citep{Sollberger2020}. Alternatively, the data can be bandpass filtered to a narrow frequency band within which the polarization properties can be assumed to be constant and the polarization states can be extracted using a time-domain approach based on the analytic signal \citep{Sollberger2018}. In this paper, we choose the S-transform \citep{Stockwell1996} to localize the polarization states of the recorded data in both time and frequency. The S-transform of the recorded data $\mathbf{\tilde{D}}(\tau, f) = [\tilde{D}_1, \ldots, \tilde{D}_6]^T(\tau, f)$ is defined as: 

\begin{equation}
\label{eq:1-9}
    \mathbf{\tilde{D}}(\tau,f)\ =\ \frac{\left|f\right|}{k\sqrt{2\pi}}
            \int_{-\infty}^{\infty}{\mathbf{\tilde{d}}(t)\exp\left(\frac{{-f}^2{(\tau-t)}^2}{2k^2}\right)}e^{-j2\pi ft}dt,
\end{equation}
where $\tau$ is the time variable (center point of the sliding Gaussian time window), $f$ is frequency, and $k$ is a scaling factor which controls the number of oscillations in the window. Increasing $k$, will increase the frequency-resolution of the S-transform at the expense of the time-resolution \citep{Schimmel2005}. 

We now form the complex covariance matrix $\mathbf{C}(\tau, f) \in \mathbb{C}^{6 \times 6}$ at each fixed point $(\tau, f)$ from the S-transformed data:

\begin{equation}
\label{eq:1-10}
    \mathbf{C}(\tau,f)=K(\Delta \tau, \Delta f)* \left(\mathbf{\tilde{D}}(\tau,\ f)\mathbf{\tilde{D}}^H(\tau,\ f)\right),
\end{equation}
where $(.)^H$ is the conjugate transpose operator and $*$ marks a 2D convolution with an averaging kernel $K(\Delta \tau, \Delta f) \in \mathbb{R}^2$, with widths in time and frequency of $\Delta \tau$ and $\Delta f$, respectively. Since seismic waves are band-limited transients, the polarization properties of pure-state modes should remain constant within a finite observation time and over a finite frequency-band. The width of the averaging kernel should be chosen accordingly. Typically, $\Delta \tau$ should be chosen such that it is frequency-dependent and proportional to $1/f$ (one dominant period) and $\Delta f$ such that it is narrow enough to ensure that dispersion effects of the estimated polarization properties are avoided. Choosing larger widths for the kernel helps to maximize the signal-to-noise ratio when extracting the polarization states but comes at the risk of potentially having multiple events with different polarization states interfere within the analysis window. Note that without the convolution with the kernel function, the covariance matrix $\mathbf{C}(\tau, f)$ would be of rank 1 at each fixed point $(\tau, f)$. The kernel $K(\Delta \tau, \Delta f)$ can be a Gaussian or a simple box function that computes a moving average of the covariance matrices over time and frequency. In the applications shown in this paper, we will use the latter.

In the following, we will drop the dependency of $\mathbf{C}$ on $(\tau, f)$ for notational brevity. We can now expand the matrix $\mathbf{C}$ as

\begin{equation}
    \label{eq:1-11}
    \mathbf{C} = \sum_{i=1}^{6}{\lambda_i \mathbf{v}_i \mathbf{v}_i^H},
\end{equation}
where $\lambda_i \in \mathbb{R}$ ($i$=1,6) are the eigenvalues and $\mathbf{v}_i \in \mathbb{C}^{6}$ are the eigenvectors of $\mathbf{C}$. By definition, $\mathbf{C}$ is Hermitian, meaning that it will have real eigenvalues and the eigenvectors form an orthonormal basis, i.e. $\mathbf{v}_j^H\mathbf{v}_k = \delta_{jk}$.

Each eigenvector $\mathbf{v}_i$ can be multiplied by a phase factor $e^{j\zeta}$ and the resulting vector will still be an eigenvector of $\mathbf{C}$. Following \citet{Samson1980}, we choose this phase factor such that the real and imaginary parts of $\mathbf{v}_i$ are orthogonal, i.e. $\Re({e^{j\zeta}\mathbf{v}_i})^T\Im(e^{j\zeta}{\mathbf{v}_i})=0$, yielding

\begin{equation}
    \label{eq:1-12}
   \zeta = -\frac{1}{2}\arctan \left(\frac{2\Re{(\mathbf{v}_i)^T \Im({\mathbf{v}_i})}}{\Re(\mathbf{v}_i)^T\Re(\mathbf{v}_i)-\Im(\mathbf{v}_i)^T\Im(\mathbf{v}_i)}\right).
\end{equation}

After the multiplication with the phase factor, the real and imaginary parts of the eigenvector correspond to the directions of the major and the minor semi-axis of the polarization ellipse, respectively.
 
If $\mathbf{C}$ contains a single pure-state wave (i.e. $n=1$ in Eq.~(\ref{eq:1-7}) within the time-frequency window), the polarization remains constant within the analysis window $K(\Delta \tau, \Delta f)$ while random noise tends to cancel out. This results in a drop of the rank of the covariance matrix, meaning that one eigenvalue $\lambda_1$ of $\mathbf{C}$ will be significantly larger than the others. All information on the polarization state of the wave will then be contained in the corresponding eigenvector $\mathbf{v}_1$. Depending on the wave type, this eigenvector will be aligned with one of the polarization vectors in Eqs.~(\ref{eq:1-2})--(\ref{eq:1-6}), after multiplication with the phase factor $e^{j\zeta}$.

If $\mathbf{C}$ contains multiple arrivals or purely random noise then more than one eigenvalue will be non-zero and the corresponding eigenvectors do not necessarily align with one of the polarization models anymore. A helpful measure to evaluate whether $\mathbf{C}$ represents a pure state is the  degree of polarization $P^2$, which can be expressed by the eigenvalues of the covariance matrix in the following way \citep[Eq. 18 in][]{Samson1980}

\begin{equation}
    \label{eq:1-13}
    P^2 =  \frac{\sum_{j=1}^{6}\sum_{k=1}^{6}{(\lambda_j-\lambda_k)^2}}{{10\left( \sum_{j=1}^6{\lambda_j}\right)^2}}.
\end{equation}
If $P^2=1$, then $\mathbf{C}$ represents a pure state and the eigenvector of the  only significant eigenvalue should align with one of the polarization vectors in Eqs~(\ref{eq:1-2})-(\ref{eq:1-6}). If $\mathbf{C}$ only contains random noise, the polarization is isotropic (meaning it does not have a preferred direction) and all eigenvalues will be non-zero and of similar order of magnitude so that $P^2\to0$.

\subsection{Efficient wave type fingerprinting by machine learning}
To infer the wave type of an extracted polarization state at a local point $(\tau, f)$, one can try to fit each of the polarization vectors in Eqs~(\ref{eq:1-2})-(\ref{eq:1-6}) to the principal eigenvector $\mathbf{v}_1$ of $\mathbf{C}(\tau, f)$ to find the one that fits best. Doing so typically requires a grid-search over all parameters $\bm{\theta}$ for each wave model $\mathbf{h}$. Various estimators can be used to evaluate the fit of a specific polarization model, for example, the maximum likelihood method \citep{Christoffersson1988, Marano2013, Sollberger2020}, orthogonal distance regression \citep{Wassermann2016}, or the MUSIC algorithm \citep{Sollberger2018}. The latter has the advantage of enabling the simultaneous estimation of the wave modes and wave parameters of multiple superimposed pure states. However, such grid-search approaches are computationally expensive due to the large parameter space that needs to be explored. Additionally, the procedure has to be repeated at each pixel $(\tau, f)$ in the time-frequency plane. This prohibits the application of such approaches to large data volumes. 

In this paper, we propose an alternative approach that separates the problem of estimating wave parameters from the classification of the wave type, thereby radically speeding up the process. To do so, we use the supervised machine learning method of support vector machines (see Appendix~\ref{svm} for details on the implementation of the SVM algorithm). A similar approach has been previously proposed  by \citet{Barak2017}, but in contrast to the approach described therein, we train our classifier in a completely data-independent fashion using a collection of theoretical 'fingerprints' given by the analytical polarization vectors in Eqs~(\ref{eq:1-2})-(\ref{eq:1-6}). The polarization vectors are parameterized with random parameters drawn from a uniform distribution. The resulting classifier is therefore fully independent of data errors and generalizes well to different data sets. This is also fundamentally different from three-component machine-learning based phase identification schemes that typically rely on large training data sets of labeled seismograms \citep[e.g.,][]{Zhu2018, Ross2018}.

To train the support vector machine (see Appendix~\ref{svm} for details), we generate a training data set consisting of $N$ polarization vectors, each with a random parametrization for each of the wave parameters $\alpha$, $\beta$, $\psi$, $\phi$, $\xi$, $c_L$, and $c_R$. Each $i$'th sample in the training data has twelve features $\mathbf{x}_i \in \mathbb{R}^{12}$ (i.e., the elements of the real and imaginary parts of the polarization vector) and a label $y_i$ corresponding to one out of a total of six classes, either one of the five wave types or noise. The noise class comprises purely random polarization vectors, meaning that all twelve features are selected from a random distribution. This additional noise class offers an alternative measure to the degree of polarization (Eq.~\ref{eq:1-13}) to evaluate whether a polarized wave is present at a specific point $(\tau, f)$ with the advantage that known characteristics of the sensor noise (e.g., the noise variance on each of the six channels) can theoretically be included in the training. After training, the classifier can be used to efficiently predict (or fingerprint) the wave type of an extracted polarization state $\mathbf{v}_1$ (the principal eigenvector of the covariance matrix). 

\begin{figure}
    \centering
    \includegraphics[width=1.0\columnwidth]{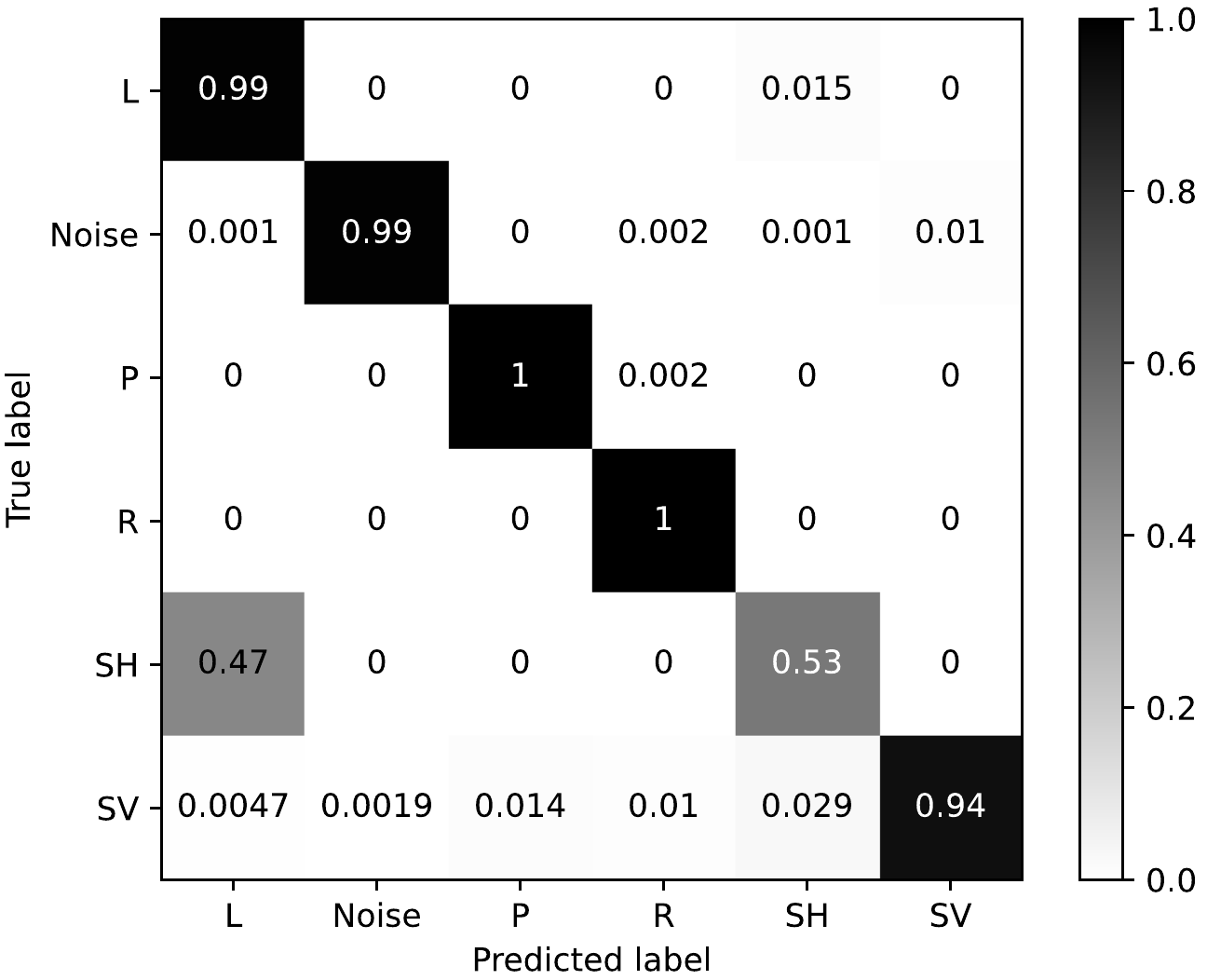}
    \caption{Evaluation of the wave type classification performance of the proposed support vector machine algorithm on an independent test data set.}
    \label{fig:confusion_matrix}
\end{figure}

\subsubsection{Classification performance}
\label{sec:classification_performance}
To evaluate the performance of the proposed classifier, we train an SVM using 5000 random polarization vectors for each class, resulting in a training data set that consists of a total of $N=6 \times 5000$ samples. We parameterize the polarization vectors using velocities that are typical for a near-surface setting. The velocities are randomly drawn from a uniform distribution on the following intervals: $\alpha \in [400$m/s$, 3000$m/s$]$, $c_R \in [100$m/s$, 3000$m/s$]$, and $c_L \in [100$m/s$, 3000$m/s$]$. The shear wave velocity $\beta$ is computed from $\alpha$ using randomly selected P-wave to S-wave velocity ratios drawn from the interval $\alpha/\beta \in [1.7, 2.4]$. We allow for the classification of waves arriving from all directions (i.e., $\phi \in [0^{\circ}, 360^{\circ}]$ and $\psi \in [0^{\circ}, 90^{\circ}]$). The Rayleigh wave particle motion can either be prograde or retrograde and we allow Rayleigh waves to be both elliptically and rectilinearly polarized (i.e., $\xi \in [-90^{\circ}, 90^{\circ}]$). 

Note that, in this example, we keep the range of parameters relatively wide as to not introduce any bias during the training. The trained classifier should therefore be able to detect wave types independent of their arrival direction and their local propagation velocity.  Any prior knowledge on the local wavefield parameters can be used in the specification of the training parameters to steer the classifier to only detect specific waves (e.g., waves arriving from a certain direction or with a certain speed) thereby achieving an even better separation of the wave types. 

After training, we evaluate the performance of the classifier on a test set of another $6000$ randomly generated polarization vectors ($1000$ samples for each class with parameters drawn from the same distributions as above) that are independent of the polarization vectors used in the training. Out of the $6000$ tested polarization vectors, $90.5\%$ are classified with the correct label, confirming that the classification problem is well separable and that 6C polarization is indeed an effective indicator of the seismic wave type, even for the very wide range of wave parameters used in the training. 

To better understand the samples of the test data that are assigned an incorrect label, we inspect the so-called confusion matrix in Fig.~\ref{fig:confusion_matrix}. The number listed in each element of the matrix indicates the percentage of samples in the test data that belong to a true class indicated by the row of the matrix and that receive a specific predicted label indicated by the column of the matrix. Ideally, if the problem was perfectly separable, only the elements on the diagonal of the matrix would be non-zero. Inspecting the confusion matrix, it becomes clear that most of the wrong predictions can be explained by a confusion between SH- and Love-waves. This is not surprising, since both waves are characterized by the same finger-print, as previously mentioned. SV-waves are assigned the wrong label in $6\%$ of the tested cases, where most confusion occurs with Love- and SH-waves. When inspecting the 6C polarization vector of SV-waves (Eq.~\ref{eq:1-3}), it becomes apparent that at the special case of close-to-vertical incidence (i.e., $\psi \to 0^{\circ}$), SV-motion becomes indistinguishable from SH-motion since the rotational motions disappear and all motion is restricted to the horizontal, translational components. The other wave types are all classified with an accuracy above $99\%$.

It should be noted that any machine learning technique could be applied for this classification task (e.g., neural networks, random forest classifiers, etc.). The choice of using a support vector machine was motivated by the fact that the algorithm only requires little training data to provide satisfactory results, is memory-efficient, and provides good classification results for the problem at hand. Additionally, it is a relatively simple algorithm seeking a linear separation of the data and thus provides more intuition in the interpretation of the results without having the 'black-box' character that is inherent to other machine learning techniques such as neural networks. 

\begin{figure*}
    \centering
    \includegraphics[width=0.7\textwidth]{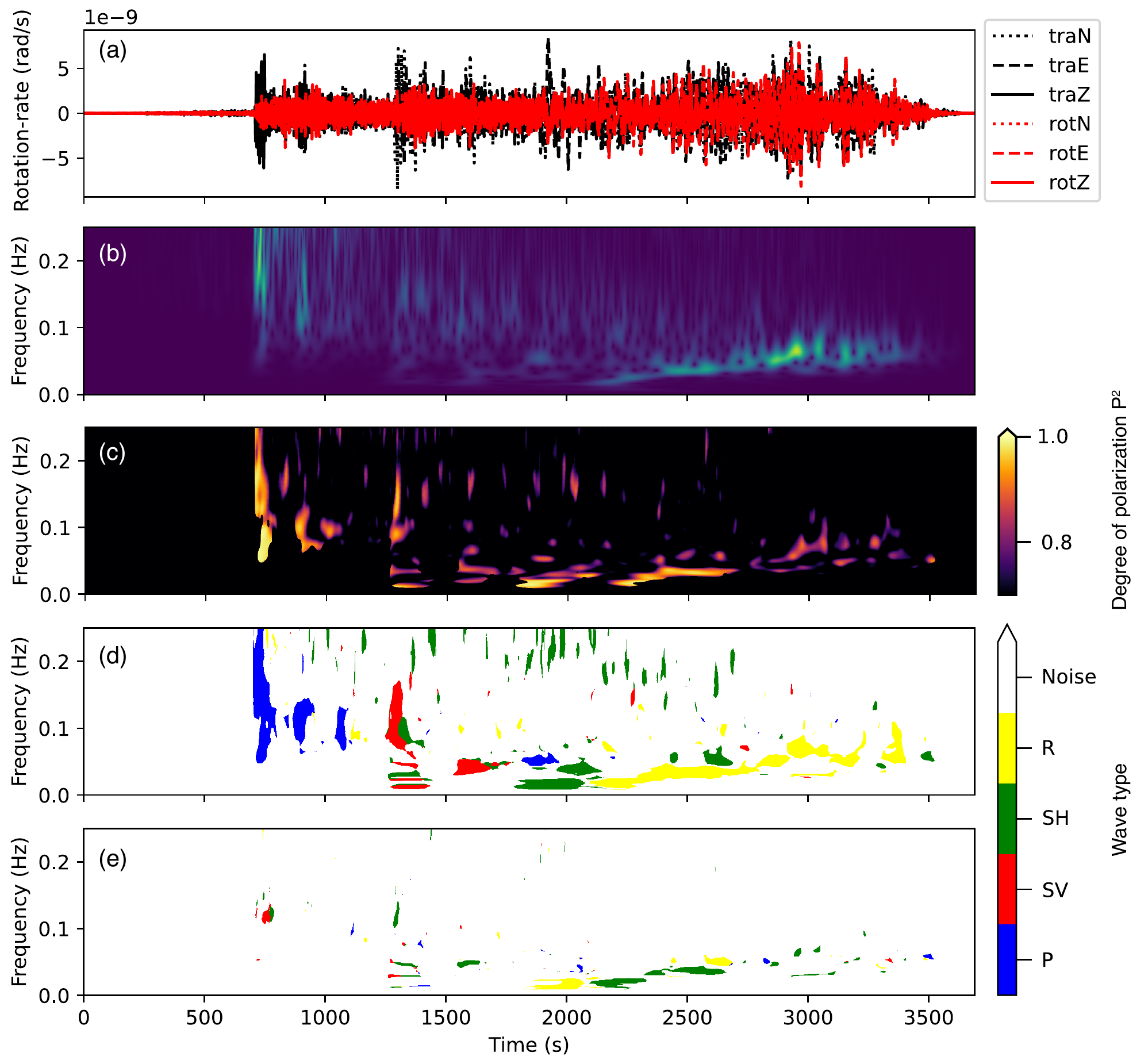}
    \caption{Example of time-frequency dependent automated seismic phase identification by 6C polarization analysis. (a) 6C seismogram of the 2018 gulf of alaska earthquake. The translational components have been converted to a 'pseudo rotation rate' (refer to the text for details). (b) S-transform of the vertical, translational component. (c) Six-component degree of polarization (colormap clipped at 0.7). (d) Wave type classification of the first eigenvector. (e) Wave type classification of the second eigenvector. For better readability, results in (c)--(e) are only plotted at time-frequency points where the signal level exceeds 5\% of the maximum amplitude.}
    \label{fig:alaska}
\end{figure*}

\subsection{Wavefield separation}
Employing the classifier described above yields a classification with labels $Y(\tau, f) \in [1, 6]$, assigning each time-frequency pixel either to one of the five wave types or to the noise class. We can now design a filtering mask $\mathbf{F}(\tau, f)=[F_1, \ldots, F_6]^T(\tau, f) \in [0, 1]$ that either isolates or suppresses specific wave modes. The mask takes on values of $1$ at points in the time-frequency plane where the desired wave mode has been detected and $0$ everywhere else. Instead of simply multiplying the filtering mask with the data at each point $(\tau, p)$, we first project the data onto a new coordinate frame that aligns with the eigenvectors $\mathbf{v}_i$ of $\mathbf{C}(\tau, f)$. The data is then multiplied with the filtering mask and subsequently projected back into the original coordinate frame to yield the S-transform of the filtered data $\mathbf{\tilde{D}}_F(\tau, f)$ as

\begin{equation}
    \label{eq:projection}
    \mathbf{\tilde{D}}_F(\tau, f) = \left(\mathbf{F}(\tau, f)\circ\left[\mathbf{\tilde{D}}(\tau, f)^T\mathbf{V}(\tau, f)\right]\right)\mathbf{V}^H,
\end{equation}
where $\mathbf{V}(\tau, f)$ is a matrix whose columns are the eigenvectors of $\mathbf{C}(\tau, f)$ and $\circ$ denotes the Hadamard product (element-wise multiplication). Depending on whether a polarization state should be kept or removed from the data, specific  elements of the filtering mask can be chosen to be non-zero. For example, if the detected wave mode at point $(\tau,f)$ should be kept in the filtered data, the filter mask is defined such that $F_1(\tau, f)=1$ and $F_{2, \ldots, 6}(\tau, f)=0$. In this way, all motion that is orthogonal to the motion of the dominant polarization state is removed. If a specific wave type should be suppressed $F_1(\tau, f)=0$ and $F_{2, \ldots, 6}(\tau, f)=1$, meaning that any underlying signal that is orthogonal to the dominant polarization state will be preserved in the filtered data. The projection onto the eigenvectors in Eq.~\ref{eq:projection} thus ensures the best possible separation with the least amount of leakage of other wave modes. 

After filtering, the inverse S-transform is applied to bring the data back to the time domain, yielding filtered 6-C waveforms $\mathbf{\tilde{d}}_F(t)\in\mathbb{R}^6$. Here, we use the inverse S-transform proposed by \citet{Schimmel2005}, providing a better time-localization of the filtered data compared to the conventional inverse S-transform
\begin{equation}
    \label{eq:ist}
    \mathbf{\tilde{d}}_F(t)=\ k\sqrt{2\pi}\int_{-\infty}^{\infty}\frac{\mathbf{\tilde{D}}_F(\tau, f)}{\left|f\right|}e^{+i2\pi ft}df.
\end{equation}
It should be noted that this inverse transform is an approximation to the true inverse S-transform, even though a very good one. The level of approximation is described by \citet{Simon2007}. Due to the better time localization of the filtered data compared to the conventional S-transform, we prefer to use this inverse transform over the conventional inverse S-transform in this application.

\begin{figure*}
    \centering
    \includegraphics[width=0.7\textwidth]{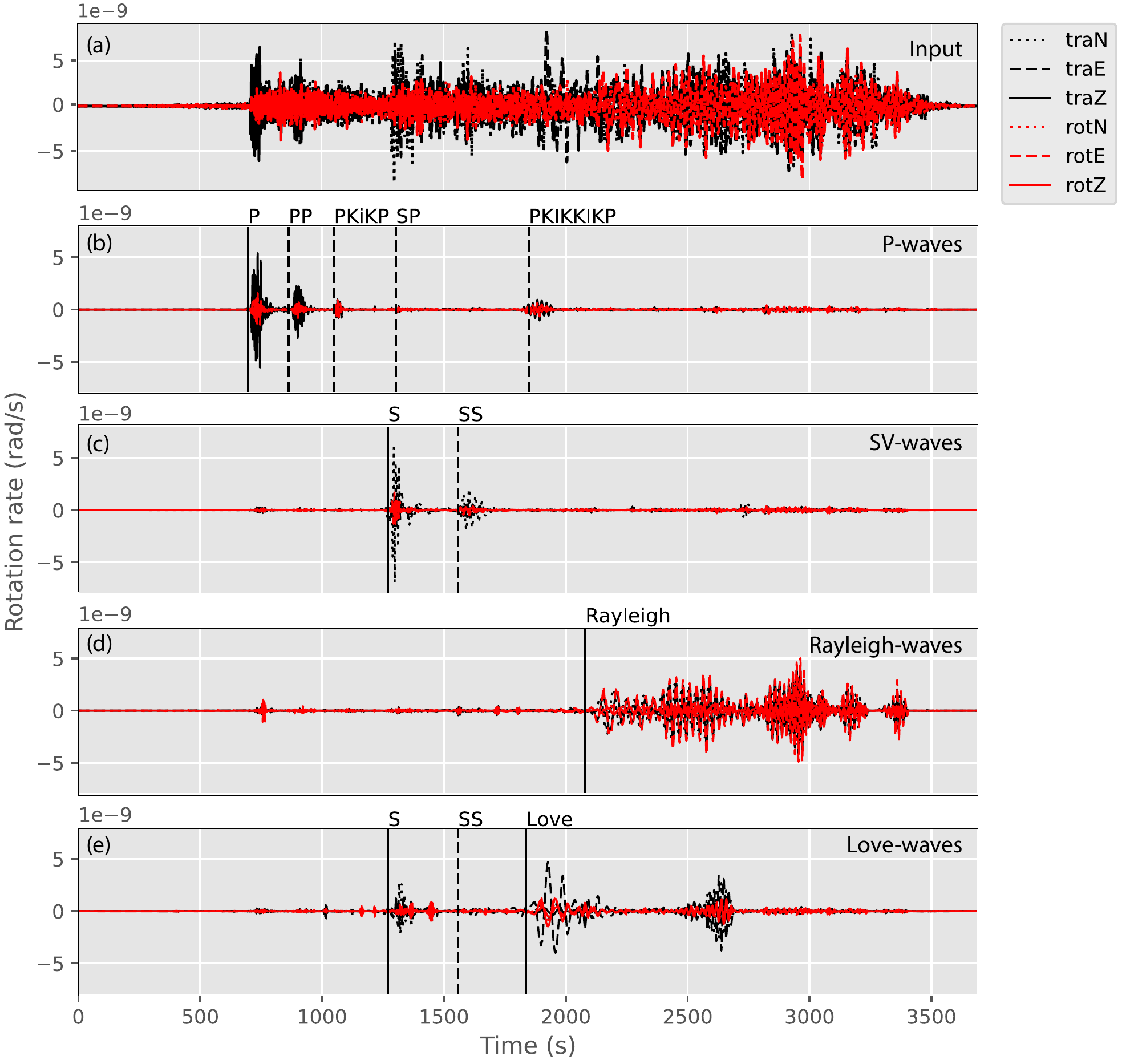}
    \caption{Wavefield separation on the example of the 2018 gulf of alaska earthquake. (a) Six-component input data and separated (b) P-wave, (c) SV-wave, (d) Rayleigh-wave, and (e) Love-wave components. Vertical lines mark predicted arrival times of some key phases obtained by ray tracing in the model of \citet{Kennett1991}. Note that the time axis is relative to the event origin time.}. 
    \label{fig:alaska_sep}
\end{figure*}

\section{Examples}
We illustrate the proposed wave type classification and wavefield separation approach with three use cases. First, the algorithm is tested on a field data example of a teleseismic earthquake recorded on the ROMY ring laser gyroscope \citep{Igel2021}. We show that the approach highly facilitates the interpretation of seismograms in the single-station case and enables the automatic picking of different seismic phases. We then apply the algorithm with the goal to identify and suppress complex, scattered surface waves in land seismic exploration data where we use a synthetic data set that is realistic for a challenging karstified environment. Ultimately, we show how the proposed algorithm can be used to efficiently detect Rayleigh- and Love-waves in ambient noise recordings, allowing for the assessment of the source distribution and source types in the noise field and the close-to real time extraction of surface wave dispersion curves, which could, in the future, prove useful for monitoring applications. 

\subsection{Single-station analysis of an earthquake recording}
We apply the proposed phase identification scheme to data recorded on the multicomponent ring-laser observatory ROMY -- located in F\"urstenfeldbruck, Germany -- constituting the most accurate six degree-of-freedom ground motion measurement system to date \citep{Igel2021}. The data was recorded after the occurrence of a magnitude 7.9 earthquake in the gulf of alaska (exact occurrence time: 2018-01-23 09:31:40 UTC). 

Fig.~\ref{fig:alaska}(a) shows one hour of 6C data as recorded by ROMY, where $t=0$~s corresponds to the origin time of the gulf of alaska earthquake. The translational components are displayed in black and the rotational components in red. The rotational data is shown in units of rotational velocity (rad/s) and the translational data was converted to the same units by applying a scaling slowness $p$ to the acceleration seismograms, where $p$ was computed according to Eq.~\ref{eq:1-8} (yielding a value of $p\approx4500^{-1}$s/m). 

The S-transform (computed with $k=1$) of the vertical, translational component is displayed in Fig.~\ref{fig:alaska}(b) showing the onset of body-wave phases (P-, and S-waves) above 0.05~Hz and the dispersive surface-wave train at lower frequencies after about $2000$~s. 

Time- and frequency-dependent six-component polarization states were computed within a time-frequency window $K(\Delta \tau, \Delta f)$ that extended over 5 dominant periods in time ($\Delta \tau=5 \times 1/f$ seconds) and over 0.1~mHz in frequency ($\Delta f=1\times10^{-4}$Hz) according to Eqs~\ref{eq:1-10}--\ref{eq:1-13}. The six-component degree of polarization (Eq.~\ref{eq:1-13}) is shown in Fig.~\ref{fig:alaska}(c) indicating regions in the time-frequency plane where pure states of motion are detected (i.e., the polarization remains constant within the specified time-frequency window). A high degree of polarization (Eq.~\ref{eq:1-13})is observed at the onset of the first-arriving motion and in specific regions in the seismic coda. 

We then applied an SVM classifier to the extracted polarization states as described above. The training of the SVM was performed with 6C polarization models parameterized randomly with samples drawn from the following uniform distributions: $\alpha, c_R, c_L \in [1000$m/s$, 4000$m/s$]$, $\alpha/\beta \in [1.7, 2.4]$,  $\phi \in [0^{\circ}, 360^{\circ}]$, $\psi \in [0^{\circ}, 80^{\circ}]$, and $\xi \in [-90^{\circ}, 90^{\circ}]$. We excluded waves at close-to-horizontal incidence from the training, i.e. $\psi > 80^{\circ}$, to avoid confusions between different wave types that can occur in this inclination angle range (see section~\ref{sec:classification_performance}). Since SH- and Love-waves cannot be distinguished from each other, we assign them to a single class comprising all SH-type motion. 

The obtained classification results with labels $Y(\tau, f)$ obtained for the principal eigenvector $\mathbf{v}_1$ are color-coded and displayed in Fig.~\ref{fig:alaska}(d). The classifier accurately assigns the first-arriving motion to the P-wave class (blue). Three distinct P-wave arrivals are identified. The onset of the direct S-wave is identified at about 1260~s, where the classification result suggest a mix of SV-type (in red) and SH-type (in green) motion. The S-wave motion is trailed by the surface wave train with Love-waves arriving first (in green) followed by highly dispersive Rayleigh waves (in yellow). 

We further evaluated, whether the classification scheme could be applied to identify multiple polarization states that overlap in both time and frequency. This should be possible as long as the polarization states in question are close-to-orthogonal to each other. This can be the case for overlapping SH- and SV-waves or Love- and Rayleigh waves arriving from similar directions. We therefore applied the classifier a second time, this time to the eigenvector $\mathbf{v}_2$ that is associated with the second largest eigenvalue. The results are displayed in Fig.~\ref{fig:alaska}(d). Most of the second eigenvectors are assigned to the noise class, meaning that a single wave mode dominates the particle motion at most points $(\tau, f)$ in the time-frequency plane. However, it can be observed that the surface wave train is composed of both Love- and Rayleigh-waves where Love-wave motion dominates over Rayleigh-wave motion in the early parts of the surface wave train and vice versa in the later parts. As expected, regions in the time-frequency plane where two overlapping wave modes are detected coincide with regions that show a relatively low degree of polarization. 

We then filtered the data according to Eq.~\ref{eq:projection} and applied the inverse S-transform (Eq.~\ref{eq:ist}) to obtain the single wave-type seismograms displayed in Fig.~\ref{fig:alaska_sep}(b)--(e). The vertical lines in Fig.~\ref{fig:alaska_sep} represent ray-theoretical arrival times predicted in the $iasp91$ model of \citet{Kennett1991}. Note that the arrival times of significant key phases in the separated data matches with the predicted arrival times obtained from ray tracing. Some Rayleigh wave energy seems to arrive before the arrival of the main Rayleigh wave train in the separated data. This could potentially be explained by near-surface P-to-Rayleigh conversions at higher frequencies, which have been previously observed for teleseismic earthquakes \citep[e.g.,][]{Zhang2022}. 

The separated seismograms are significantly easier to interpret than the original six-component data (Fig.~\ref{fig:alaska_sep}a) and could be used, for example, for automated picking of specific phase arrivals, for pick refinements, or for the identification of exotic phases that are hard to be found in the raw data. Additionally, the separated data could provide a way to compute more reliable receiver functions.

\subsection{Aliased ground-roll suppression}
\begin{figure}
    \centering
    \includegraphics[width=1.0\columnwidth]{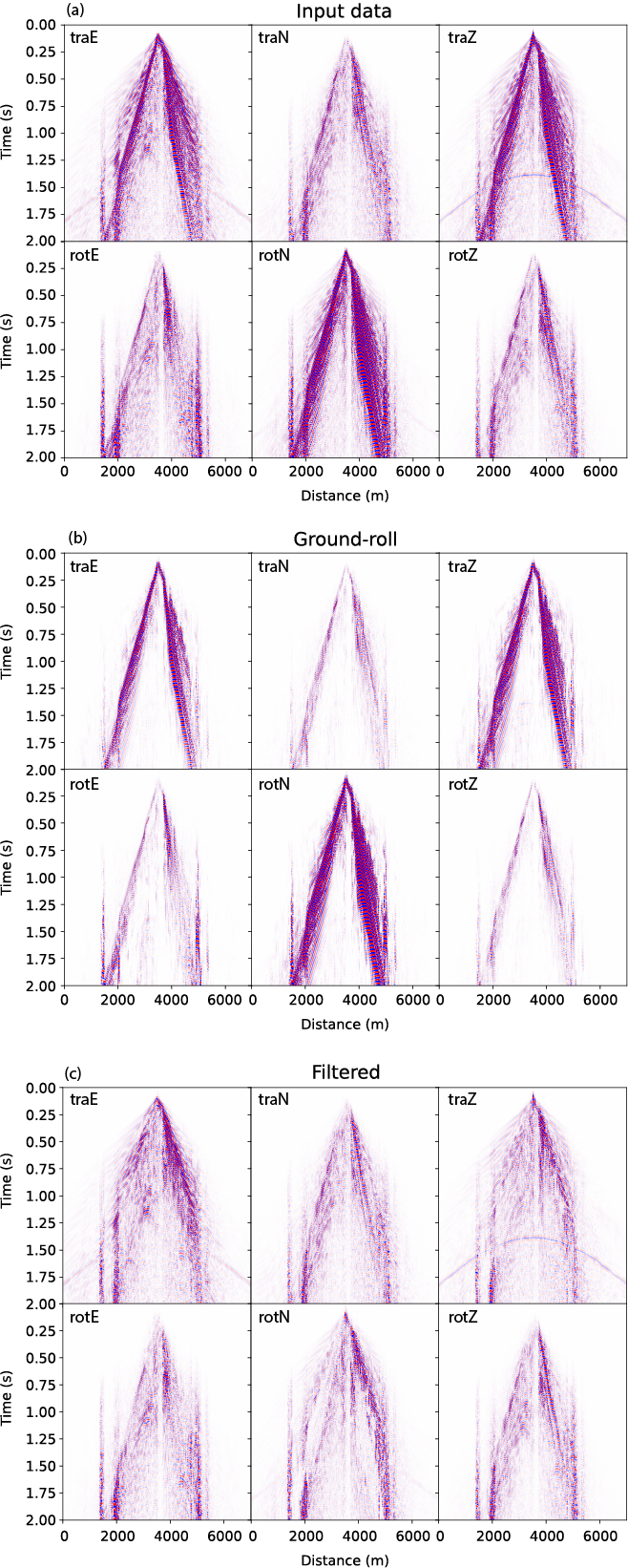}
    \caption{Suppression of aliased and side-scattered ground-roll using the proposed filtering method. (a) Six-component shot gather used as an input for filtering. (b) Separated ground-roll component (Love- and Rayleigh waves) obtained individually at each recording position. (c) Remaining signal after the suppression of ground-roll.}.
    \label{fig:ground-roll}
\end{figure}

Surface waves (often referred to as ground-roll) pose a major challenge to land seismic exploration since they can obscure reflection signals from subsurface reflectors of interest. Conventional ground-roll suppression techniques such as f-k- or $\tau$-p-filtering rely on the identification and isolation of the surface waves based on their lower moveout velocity across receiver arrays compared to body waves and require that the surface waves are appropriately sampled in space. Since ground-roll is the shortest wavelength component in the recorded data, this undesired signal component typically dictates the spatial sampling in land seismic exploration. Additionally, conventional ground-roll filtering techniques struggle with the suppression of side-scattered surface waves. 

As an alternative to array processing, ground roll can be identified and suppressed with a single geophone by the application of three-component polarization filtering \citep[e.g.,][]{DeFranco2001, Jin2005, Kendall2006}, enabling the filtering of spatially aliased ground-roll with the potential to significantly lower the field effort and thus the survey costs. This kind of filtering is based on the assumption that ground-roll purely consists of elliptically-polarized Rayleigh waves. In reality, this assumption is not necessarily fulfilled since ground-roll can sometimes retain a linear polarization \citep{Kragh1995} and the horizontal geophone components can be significantly contaminated by Love waves. 

Rotational data can help to improve single-receiver ground-roll identification and suppression \citep[e.g.,][]{Muyzert2012, Edme2013, Barak2014, Barak2017, Muyzert2019, Allouche2020} since they provide an additional, powerful discrimination criterion besides the ellipticity to distinguish body-waves from ground-roll:  namely, the local propagation velocity of the wave. Since the amplitudes of rotational motions inversely scale with the wave's velocity (see for example factor $c_R^{-1}$ in Eq.~\ref{eq:1-5}), rotational data is typically dominated by the low-velocity ground-roll, thereby facilitating its identification and suppression. 

Here, we test the proposed machine-learning based wavefield separation scheme for the single-station suppression of aliased ground-roll. Since rotational sensors are not yet widely available for use in land-seismic exploration, we test the scheme on a synthetic data set obtained via elastic finite-difference simulations in the Arid model of the SEAM Phase II consortium for land-seismic challenges \citep{Oristaglio2015}. The Arid model was designed to represent an extreme desert environment with geological features such as karsts, wadis, sand, outcropping bedrock and unusual topography. All of these features introduce anisotropy and can cause intense scattering of elastic waves and in the near-surface generating complex, side-scattered and aliased ground-roll that is difficult to suppress with conventional methods.

A shot gather of simulated six-component data is shown in Fig.~\ref{fig:ground-roll}(a), for an East-West oriented receiver line. To facilitate the evaluation of the wavefield separation scheme, we added a single, artificial P-wave reflection to the shotgathers (visible intercepting the time axis at about 1.75 seconds). Note that there is significant energy in the crossline North component of the translational data (labeled with traN), suggesting the presence of side-scattered waves and SH-waves. 

We applied the proposed six-component wavefield separation scheme with the goal to identify and suppress ground-roll (Rayleigh- and Love waves) using a support vector machine that was trained with the following parameter distributions: $c_R, c_L \in [400$m/s$, 1000$m/s$]$, $\alpha \in [1050$m/s$, 5000$m/s$]$, $\alpha/\beta \in [1.7, 2.4]$,  $\phi \in [0^{\circ}, 360^{\circ}]$, $\psi \in [0^{\circ}, 80^{\circ}]$, and $\xi \in [-90^{\circ}, 90^{\circ}]$. Note that the training data implicitly includes the wave velocity as a discrimination criterion. Constraints on the wave types from prior knowledge (here, the lower velocity of ground-roll compared to body waves), can thus be easily included in the filtering by restricting the parameter range of the polarization models during the training.

The result of separating the ground-roll from any underlying signal is shown in Figs~\ref{fig:ground-roll}(b) and (c). The complex ground-roll seems to be accurately detected and isolated from the data, independent of its arrival direction, resulting in a ground-roll reduction of approximately 20~dB on the vertical translational component. Since the filtering is applied locally at each receiver location, it is not affected by spatial aliasing. Note that the artificial P-wave reflection is preserved in the filtered data in Fig.~\ref{fig:ground-roll}(c).

\begin{figure*}
    \centering
    \includegraphics[width=0.8\textwidth]{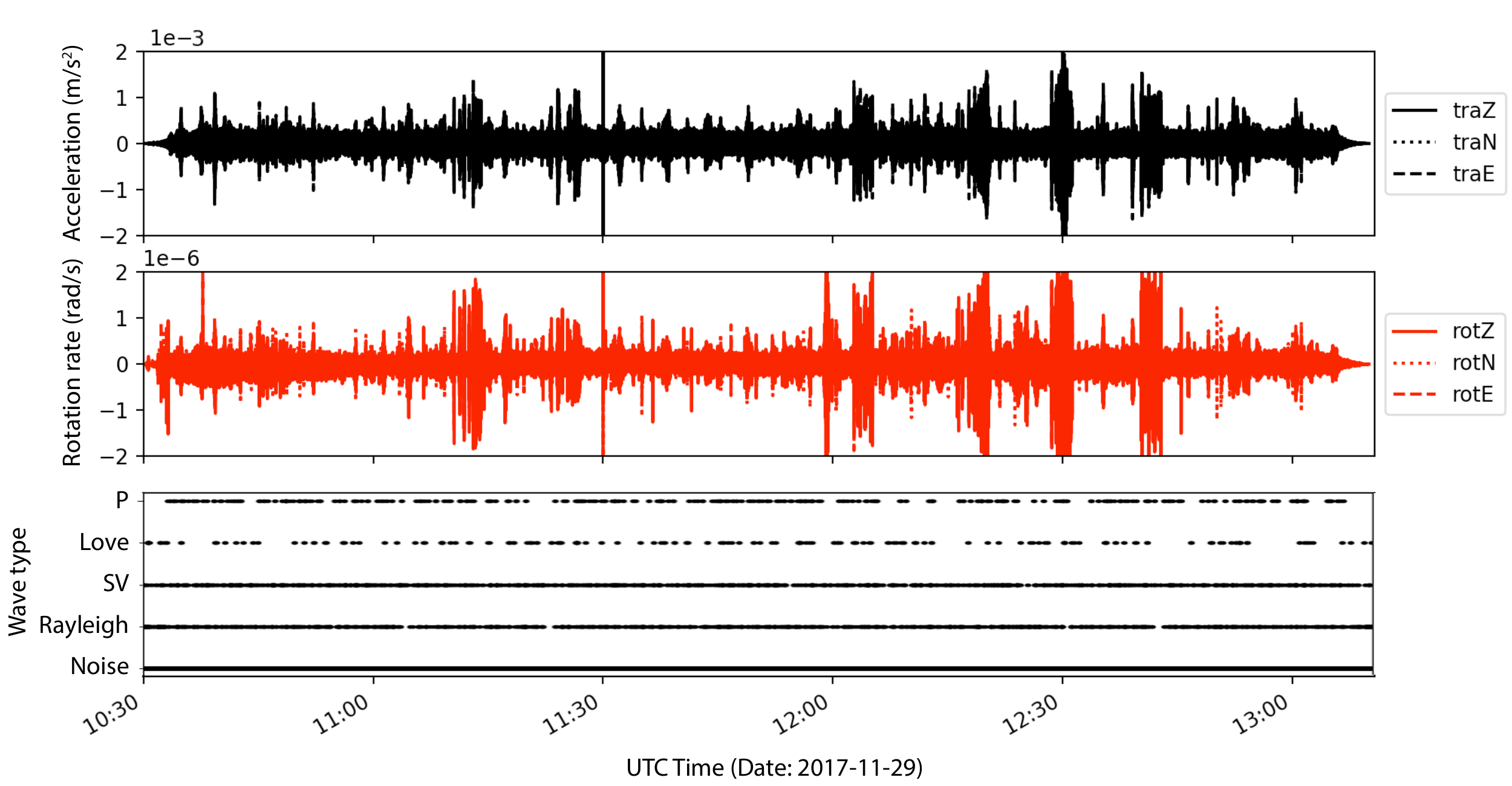}
    \caption{Six-component recording of about 2 hours and 40 minutes of urban noise. Ground acceleration is shown in black, ground rotation rate in red. The bottom panel shows the result of applying the proposed wave type classification scheme to the recordings above (classification obtained at 14 Hz). Black dots mark the detection of a specific wave type in the analysis time window. Note that the results seem to imply that the recorded wavefield is dominated by SV- and Rayleigh-wave motion.}.
    \label{fig:noise}
\end{figure*}

\subsection{Ambient noise analysis}
The analysis of rotational motions in the ambient noise wavefield can help to characterize the noise sources \citep[e.g.,][]{Hadziioannou2012, Tanimoto2015, Tanimoto2016} and to extract local surface wave dispersion curves \citep[e.g.,][]{Kurrle2010, Wassermann2016, Edme2016, Yoshida2020}.

With the recent availability of sensitive, portable rotation sensors, \citet{Keil2020} could successfully extract near-surface dispersion curves from six-component recordings of urban noise using the pioneering method of \citet{Wassermann2016}. The single-station, six-component method appears to yield comparable results to conventional array methods \citep{Keil2022}. Key advantages of the six-component approach over array methods is the overall reduced logistical effort during deployment and maintenance. Additionally, the single-station approach yields dispersion curves that are truly local, without smearing out the extracted velocity information over the aperture of an array \citep[e.g.,][]{Tang2022}. 

Here, we apply the proposed wave type fingerprinting scheme to a six-component seismic recording of urban noise, recorded in the city center of Munich with an iXblue blueSeis-3A rotation sensor together  with  a  Nanometrics Trillium Compact seismometer \citep[please refer to][for details on this dataset]{Keil2020}. We analyse about 2 hours and 40 minutes of data. The six-component recording is shown in Fig~\ref{fig:noise}. We first filtered the data with a series of narrow bandpass filters (each a quarter octave wide) to non-overlapping frequency bands between 1~Hz and 20~Hz. For each filtered time-series we then computed the data covariance matrix of the complex analytic signal within a sliding time-window that extended over two periods according to Eq.~(\ref{eq:1-10}). The time domain approach with the analytic signal was chosen over the S-transform for this application simply because it is more memory-efficient when processing such long time series. We then classified the principal eigenvector with the proposed fingerprinting scheme. The training parameters of the SVM corresponded to the ones described in Section~\ref{sec:classification_performance}. The bottom panel in Fig.~\ref{fig:noise} shows the classification result at each position of the sliding time window (for the data filtered with a bandpass filter centered at 14~Hz). The recorded wavefield appears to be mainly dominated by Rayleigh- and SV-type motion, with additional less frequent detections of P- and Love-waves.

Once the wave type is known, specific Rayleigh-wave and Love-wave parameters (phase velocity, azimuth, and ellipticity angle) can be easily extracted from their polarization states (principal eigenvectors). According to the Love-wave polarization model in Eq.~(\ref{eq:1-6}), the Love wave phase velocity for a time window centered at $\tau$ with the data filtered to the frequency $f$ can be estimated from the principal eigenvectors $\mathbf{v}_1(\tau, f)\in \mathbb{C}^6=[v_1,\ldots,v_6]^T(\tau, f)$ that were assigned to the Love wave class as
\begin{equation}
    \label{eq:1-16}
    c_L = - \frac{\left[\sin\phi_L\Re(v_1) - \cos\phi_L\Re(v_2)\right]}{2 \Re(v_6)}, 
\end{equation}
where the dependency on $(\tau, f)$ is implicit. The propagation azimuth of the Love wave is obtained by
\begin{equation}
    \label{eq:1-17}
    \phi_L = \arctantwo\left(\Re(v_2), \Re(v_1)\right) + \frac{\pi}{2}.
\end{equation}

For eigenvectors that were assigned to the Rayleigh wave class, we can estimate the Rayleigh wave phase velocity as
\begin{equation}
    \label{eq:1-18}
    c_R = \frac{\Re(v_3)}{\left[\sin\phi_R \Re(v_4) - \cos\phi_R \Re(v_5)\right]}, 
\end{equation}
where the Rayleigh wave propagation azimuth is obtained via
\begin{equation}
\label{eq:1-19}
    \phi_R = \arctantwo\left(\Im(v_2), \Im(v_1)\right).
\end{equation}
Ultimately, we can estimate the Rayleigh wave ellipticity angle as
\begin{equation}
\label{eq:1-20}
    \xi = \arctan\left(\frac{\cos\phi_R\Im(v_1) + \sin\phi_R\Im(v_2)}{\Re(v_3)}\right).
\end{equation}

In Eqs~(\ref{eq:1-17}), (\ref{eq:1-19}), and (\ref{eq:1-20}), special care needs to be given to the sign of the individual components of the principal eigenvector, to ensure that the back-azimuth can be recovered on the full 360$^{\circ}$ interval and that the correct sign of the  ellipticty angle (pro-grade or retro-grade motion) is retrieved.

Note that without the initial step of wave type classification, there would be little meaning to the values obtained by Eqs~(\ref{eq:1-16})--(\ref{eq:1-20}). The equations only yield the desired wave properties if the extracted eigenvector indeed corresponds to the polarization of the specified wave type. We therefore argue that an initial step of wave-type classification is unavoidable before estimating any wavefield properties from rotational recordings (see Section~\ref{discussion} for a more detailed discussion).



\begin{figure*}
    \centering
    \includegraphics[width=0.9\textwidth]{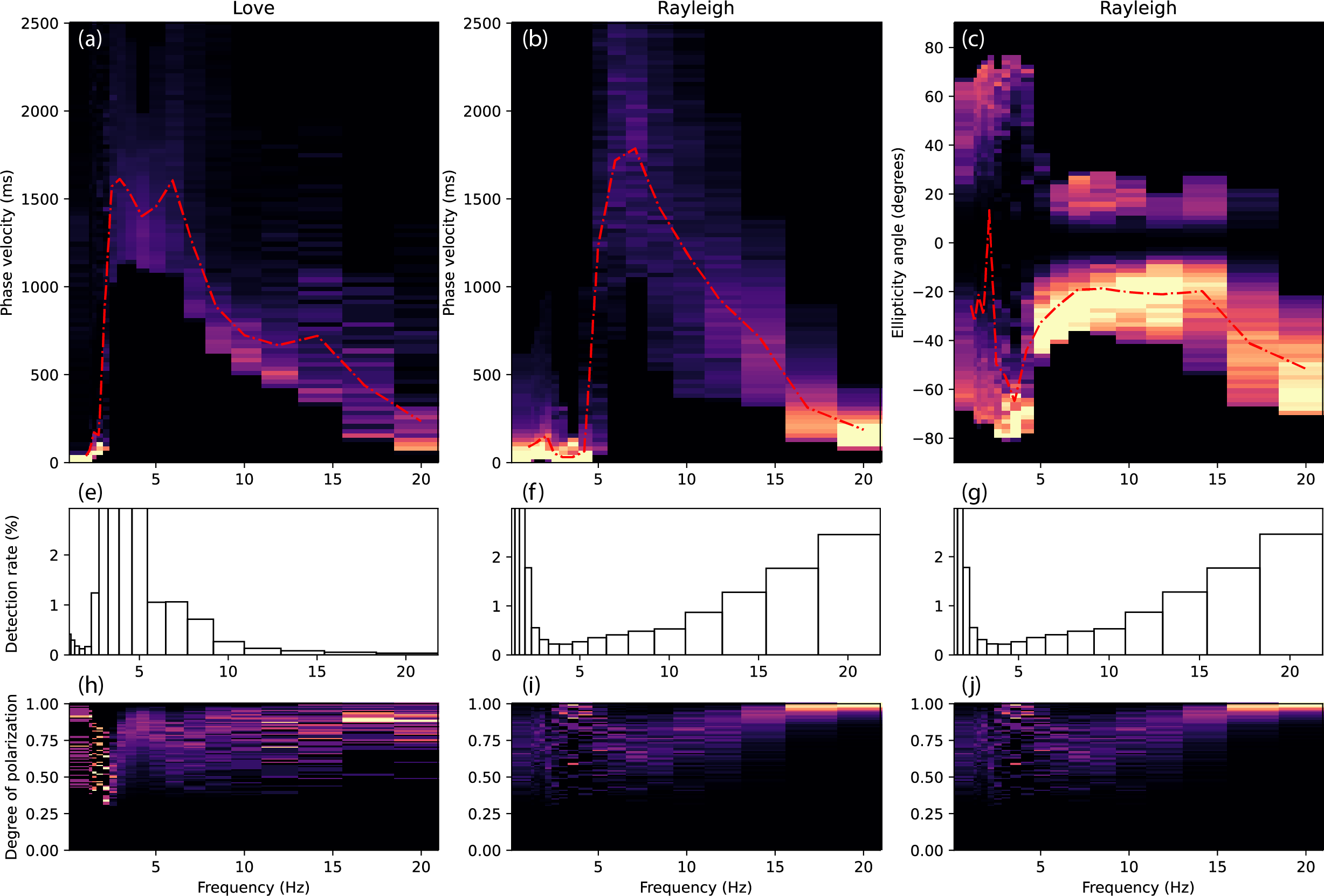}
    \caption{Frequency-dependent phase velocity for Love- (a) and Rayleigh-waves (b) and Rayleigh wave ellipticiy angle (c) estimated from the single-station six-component urban noise recording in Fig.~\ref{fig:noise}. Red lines mark the median of the parameters extracted at each frequency. Panel (e) shows, for each analysed frequency band, the percentage of the total number of analyzed time windows within which a Love wave was detected. Panels (f) and (g) are identical and show the percentage of the total number of analyzed time windows within wich a Rayleigh wave was detected. Panels (h)--(j) show the binned degree of polarization of all time windows that were assigned to the Love-wave or Rayleigh-wave class}.
    \label{fig:dispersion}
\end{figure*}

 Histograms of the estimated frequency-dependent Love- and Rayleigh-wave phase velocities and the Rayleigh-wave ellipticity angle are displayed in Fig.~\ref{fig:dispersion} for bins with a width of 25~m/s and 1.8$^{\circ}$, respectively. Both the Love- and Rayleigh-wave phase velocity show the typical dispersion characteristic with a decrease in the estimated phase velocity with frequency. Below 5 Hz, the phase velocity curve drops to 0 since the urban noise level at frequencies below 5~Hz falls below the self-noise of the rotational seismometer, as suggested by \citet{Keil2020}. As a result, below 5~Hz, the denominator in Eq.~(\ref{eq:1-16}) and Eq.~(\ref{eq:1-18}) dominates over the numerator due to the self-noise in the rotational components, leading to estimated phase velocities that are close to 0. 
 
 The estimated, frequency-dependent Rayleigh wave ellipticity angle is predominantly negative (Fig~\ref{fig:dispersion}c), indicating retrograde motion. The bifurcation of the ellipticity curve between 5~Hz and 15~Hz could indicate the presence of higher modes, which are difficult to characterize with the given approach since it is assumed that only a single wave mode is present in each analysis window. Note that the information on the sense of rotation of Rayleigh waves is not available from conventional H/V analysis with single-station three-component seismometers and can potentially help to better constrain the inversion for structural models \citep{Hobiger2016, Marano2017}.
 
 The detection rate of Love- and Rayleigh-waves at each frequency is displayed in Figs~\ref{fig:dispersion}(e)--(g), where the detection rate is defined as the percentage of the total number of analyzed time windows, within which a specific wave type was detected. This provides an intuitive way to understand the frequency-dependent wave type composition of the recorded noise field. Above 10~Hz, Rayleigh-waves seem to dominate over Love-waves in the analysed urban noise time series. Below 10~Hz, more Love- than Rayleigh-wave energy appears to be present. Spurious Love- and Rayleigh-wave detections seem to occur at frequencies below 5~Hz, where the urban noise level drops below the instrument self-noise of the rotational seismometer, suggesting that the polarization of the sensor self noise is aligned with the polarization models of Love- and Rayleigh-waves. Such misdetections could potentially be avoided if the self-noise characteristics of the used seismometers would be included in the training of the noise class of the support vector machine. 
 
 To provide an additional measure of the quality of the extracted dispersion curves, we analyse the degree of polarization (Eq.~\ref{eq:1-13}) of the six-component motion in each of the analysis time windows where a Love- or Rayleigh wave was detected (Fig~\ref{fig:dispersion}h--j). Ideally, the wave parameters are extracted from pure states with a degree of polarization close to unity. Lower values for the degree of polarization indicate that the recorded motion deviates from a pure state either due to the interference of other wave types or due to high noise levels. Therefore, less confidence should be given to estimates obtained from time windows that show a low degree of polarization. For the analysed data, a high degree of polarization close to 1 is only obtained for high frequencies ($>$15~Hz). Below that, the degree of polarization decreases with frequency both for Love- and Rayleigh-waves. We expect that this is likely due to the decrease of the signal-to-noise ratio as the signal levels approach the instrument self-noise due to the lack of strong low-frequency sources. Before inverting the dispersion curves for subsurface structure, the degree of polarization could potentially be used to weight the extracted phase velocity and ellipticity angle values to only include values with a high level of confidence (similarly to the weighting scheme proposed by \citet{Wassermann2016}). However, the inversion of the extracted dispersion curves goes beyond the scope of this paper.
 
 Due to the low computational costs of the proposed wave type fingerprinting scheme, dispersion curves such as the ones shown in Fig.~\ref{fig:dispersion} can be extracted almost in real time and from very short time series (given that Love- and Rayleigh waves are excited at the frequency-band of interest) without the need to compute cross-correlations. This could prove useful in the future for real-time monitoring applications with ambient noise. 

\section{Discussion}
\label{discussion}
Wave type fingerprinting as described in this paper is achieved using a machine-learning based classification model that is trained using analytically derived polarization models. Theoretically, these models are only valid for plane waves in an isotropic medium. The plane-wave assumption might not be fulfilled when the recording station is located close to the source. However, we expect that the violation of the plane wave assumption only has a negligible effect on the wave type classification performance, since the polarization models mainly rely on the propagation direction of the wave, which is always perpendicular to the wavefront at a single point of observation in space, even if the wavefront is not planar. Yet, it has to be noted that a violation of the plane wave assumption will have an impact on the subsequent estimation of wave parameters such as the phase velocity \citep[e.g.,][]{Wassermann2016}. In the future, it will be necessary to investigate whether the proposed scheme is applicable to data in the presence of anisotropy, where the 6C polarization can differ from the polarization of the isotropic case \citep[][]{Pham2010, Noe2022}. 

Multiple wave types that overlap in frequency and time can be classified with the proposed scheme, as long as their polarization is close to orthogonal. We expect, that the scheme can be extended to accurately fingerprint arbitrary mixtures of waves, when the elements of the null space of the covariance matrix (spanned by the eigenvectors corresponding the smallest eigenvalues) are used as features to train the classifier instead of the elements of the principal eigenvector. The null-space will always be orthogonal to the polarization states present in the covariance matrix, even if the polarization states themselves are not orthogonal. Null-space algorithms, such as the MUSIC algorithm, have been successfully used in the past to characterize multiple interfering six-component polarization states \citep{Sollberger2018}.

The extraction of phase velocities from combined measurements of rotational and translational motion typically relies on the computation of simple amplitude ratios between different components of the recorded motion. The velocities that are extracted in such a way can only provide meaningful information on the actual phase velocity, if the wave type of the recorded motion is known. Otherwise, the extracted amplitude ratios only have little meaning and can take on random values that can lead to a wrong interpretation (e.g., if an analysis time window only contains random noise). In previously existing methods, the estimation of the wave type has been implicitly included in the estimation of the wave parameters. For example, for Love waves, the estimation of the wave type is simple, since Love waves are the only wave type that exhibit rotational motion around the vertical axis. One only needs to be make sure that a Love wave is present in the analysis window before estimating the phase velocity, for example, by verifying that a linear correlation exists between the transverse translation and the vertical rotation components within the time window \citep[e.g.,][]{Hadziioannou2012, Wassermann2016}. Such approaches constitute simple wave type detection schemes with a simultaneous estimation of the wave parameters (back-azimuth and phase velocity). 

However, for other wave types, such as Rayleigh waves, such a simplified approach might fail. For example, the Rayleigh wave phase velocity estimation schemes proposed by \citet{Edme2016} or \citet{Keil2020} relies on finding a linear relation between the vertical translation  and the transverse rotation component, for example, by orthogonal distance regression for the unknown wave parameters (back-azimuth and phase velocity). We expect that this can introduce errors in the extracted dispersion curves since both P- and SV-waves are also characterized by a linear relationship between the vertical translation and the transverse rotation component according to the polarization models presented in Eq.~(\ref{eq:1-2}) and Eq.~(\ref{eq:1-3}). This can potentially lead to the leakage of body waves into the Rayleigh wave dispersion curves and possibly to an erroneous estimation of the phase velocities. This effect is probably negligible in most cases since the ambient noise field is typically dominated by surface waves, but could become significant in the presence of strong sources that radiate a significant amount of body wave energy. Here, more sophisticated wave type classification schemes, such as the one described in this paper can potentially help to avoid this issue, since it allows one to fully discriminate between P-, SV-, SH- and Rayleigh wave motion. The initial step of classification thereby ensures that minimal leakage of other wave types occurs that could contaminate the extracted wave parameters and dispersion curves.

\section{Conclusion}
We have introduced an efficient machine-learning based wave type fingerprinting and wavefield separation scheme for six-component ground-motion recordings that can be used to rapidly classify seismic phases for large volumes of single-station data. Unlike three-component recordings, six-component recordings of combined translational and rotational motion enable the full discrimination between P-, SV-, SH- and Rayleigh wave motion. 

We have shown that the proposed scheme can highly facilitate the interpretation of seismic data from sparse sensor networks and enable improved single-receiver polarization filtering to suppress ground-roll in land seismic exploration. Additionally, the proposed scheme can provide valuable insights into the wave type composition of the ambient noise field.  

After the classification of the wave type of a recorded seismic phase, wave parameters such as the phase velocity, propagation direction and ellipticity can be directly estimated from its six-component polarization state without the need for computationally expensive grid-search algorithms. We have shown that this enables the extraction of Rayleigh- and Love-wave dispersion curves and frequency-dependent Rayleigh wave ellipticity from short recordings of ambient noise almost in real-time and at low computational costs, rendering the scheme attractive for use in monitoring applications. The extracted dispersion curves are truly local without the effect of smearing out the velocity information over the aperture of an array as in conventional approaches. 

\begin{acknowledgments}
We would like to thank Heiner Igel, Joachim Wassermann, and Sabrina Keil from LMU Munich for providing the data that was used in this paper and for inspiring discussions. 

We acknowledge financial support by TOTAL Energies. D.S. received additional funding from the European Union‘s Horizon 2020 research and innovation program under grant agreement number 821881 (PIONEERS).
\end{acknowledgments}

\begin{dataavailability}
The open source software package TwistPy (\textbf{T}oolbox for \textbf{W}avefield \textbf{I}nertial \textbf{S}ensing \textbf{T}echniques) is published together with this paper at \url{https://twistpy.org}, the source code can be found at \url{https://github.com/solldavid/TwistPy}. The repository includes all the field data and Jupyter notebooks that allow the reader to fully reproduce the processing examples from this paper. Note that the synthetic SEAM phase II Arid dataset used to generate Figure~\ref{fig:ground-roll} cannot be shared due to copyright issues.
\end{dataavailability}


\bibliography{library.bib}

\appendix
\section{Amplitudes of reflected body waves at a free surface}
\label{polarisationmodel}

The free surface 6C polarization vectors in Eqs.~(\ref{eq:1-2}) and (\ref{eq:1-3}) for P- and SV-waves depend on the amplitudes of the incident and reflected waves. Here, we give explicit expressions of the amplitudes of the reflected waves as a function of the inclination angle of the incident wave $\psi$ and the local P-wave and S-wave velocties $\alpha$ and $\beta$. 

In the case of an incident P-wave of amplitude $A_P$ and inclination $\psi$, the amplitudes of the reflected P-wave $A_{PP}$ and the reflected S-wave $A_{PS}$, normalized by the amplitude of the incident wave, are given by \citep[e.g.,][]{Achenbach1973}:

\begin{equation} 
 \label{eq:14}
\frac{{{A _{PP}}}}{{{A _P}}}(\alpha, \beta, \psi) = \frac{{\sin 2{\psi}\sin 2{\psi_S} - {\kappa ^2}{{\cos }^2}2{\psi_S}}}{{\sin 2{\psi}\sin 2{\psi_S} + {\kappa ^2}{{\cos }^2}2{\psi_S}}}, 
\end{equation}

and

\begin{equation} 
 \label{eq:15}
\frac{{{A _{PS}}}}{{{A _P}}}(\alpha, \beta, \psi) = \frac{{2\kappa \sin 2{\psi}\cos 2{\psi_S}}}{{\sin 2{\psi}\sin 2{\psi_S} + {\kappa ^2}{{\cos }^2}2{\psi_S}}}.
\end{equation}
 
where the inclination angle of the reflected S-wave $\psi_S$ is given by

\begin{equation} 
 \label{eq:16}
\sin \psi_S=\kappa^{-1}\sin \psi,
\end{equation}

where $\kappa$ is the local P-wave to S-wave velocity ratio:

\begin{equation} 
 \label{eq:17}
\kappa  = {\alpha / \beta}.
\end{equation}

In the case of an incident SV-wave of amplitude $A_S$ and inclination $\psi$, the amplitudes $A_{SS}$ of the reflected S-wave and $A_{SP}$ of the reflected P-wave, normalized by the amplitude of the incident wave, are:

\begin{equation} 
 \label{eq:26}
\frac{{{A _{SS}}}}{{{A _S}}}(\alpha, \beta, \psi) = \frac{{\sin 2{\psi}\sin 2{\psi_P} - {\kappa ^2}{{\cos }^2}2{\psi}}}{{\sin 2{\psi}\sin 2{\psi_P} + {\kappa ^2}{{\cos }^2}2{\psi}}}, 
\end{equation}

and

\begin{equation} 
 \label{eq:27}
\frac{{{A _{SP}}}}{{{A_S}}}(\alpha, \beta, \psi) = \frac{{-\kappa \sin 4{\psi}}}{{\sin 2{\psi}\sin 2{\psi_P} + {\kappa ^2}{{\cos }^2}2{\psi}}},
\end{equation}

with the inclination $\psi_P$ of the reflected P-wave:

\begin{equation}
    \sin{\psi_P}=\kappa\sin{\psi}.
\end{equation}

For the special case of an incident SV-wave at the critical inclination angle $\psi=\sin^{-1}(1/\kappa)\equiv \psi_{cr}$, $\psi_P$ becomes $\pi/2$ and the amplitudes of the reflected waves are:
\begin{equation}
     \frac{A_{SS}}{A_{S}}=-1
\end{equation}
and 
\begin{equation}
    \frac{A_{SP}}{A_S}(\alpha, \beta)=[4(\kappa^2 -1)]/[\kappa(2-\kappa^2)],
\end{equation}

 Above the critical angle ($\psi > \psi_{cr}$), $\sin \psi_P>1$, and thus $\psi_P$ becomes imaginary. Quantities $\frac{A_{SS}}{A_{S}}$, and $\frac{A_{SP}}{A_S}$ then become \citep{Nuttli1961}:

\begin{equation} 
 \label{eq:Ass_crit}
\frac{{{A_{SS}}}}{{{A_S}}}(\alpha, \beta, \psi) = \dfrac{\splitdfrac{4({{\sin }^2}{\psi} - {\kappa ^{ - 2}}){{\sin }^2}2{\psi}{{\sin }^2}{\psi} - {{\cos }^4}{\psi}} {+ 4j\sqrt {{{\sin }^2}{\psi} - {\kappa ^{ - 2}}} \sin 2{\psi}\sin {\psi}{{\cos }^2}2{\psi}}}{{{{\cos }^4}2{\psi} + 4({{\sin }^2}{\psi} - {\kappa ^{ - 2}}){{\sin }^2}2{\psi}{{\sin }^2}{\psi}}},
\end{equation} 

and

\begin{equation} 
 \label{eq:Asp_crit}
\frac{{{A_{SP}}}}{{{A_S}}}(\alpha, \beta, \psi) = \dfrac{{\splitdfrac{2{\kappa ^{ - 1}}\sin 2{\psi}\cos 2{\psi}}{({{\cos }^2}2{\psi} - 2j\sqrt {{{\sin }^2}{\psi} - {\kappa ^{ - 2}}} \sin 2{\psi}\sin {\psi})}}}{{{{\cos }^4}2{\psi} + 4({{\sin }^2}{\psi} - {\kappa ^{ - 2}}){{\sin }^2}2{\psi}.{{\sin }^2}{\psi}}}.
\end{equation} 

Here, $j$ indicates that SV particle motion above the critical angle is not rectilinear anymore, but elliptical.

\section{Support vector machine classification}
\label{svm}
      
\begin{figure}
    \centering
        \includegraphics[width=0.7\columnwidth]{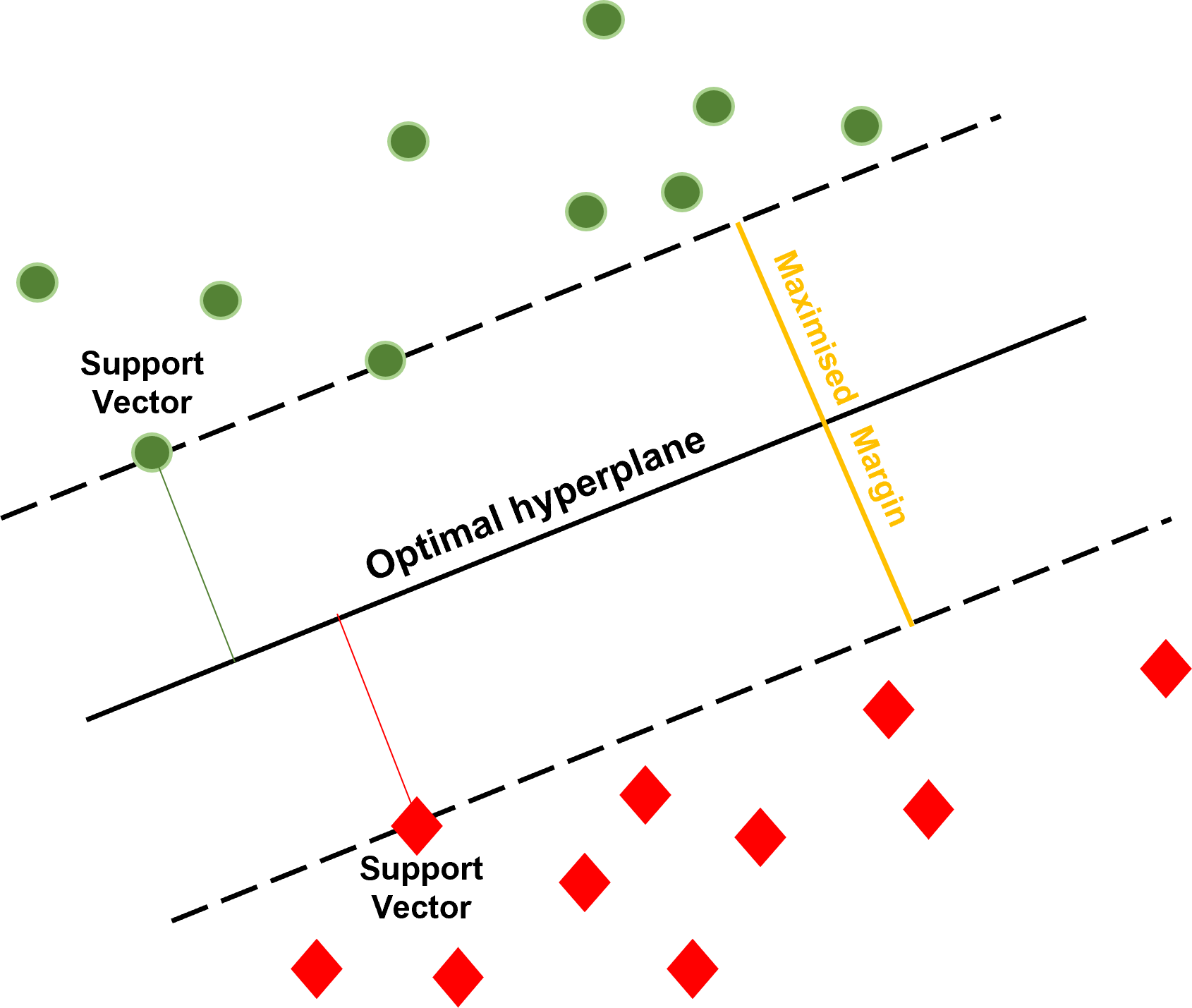}
    \caption{Schematic illustration of Support Vector Machine (SVM) approach in the case of linearly separable data in a two-dimensional feature space (i.e., the separating hyperplane becomes a line).} 
    \label{Figure 3-1}
\end{figure}

The Support Vector Machine (SVM) algorithm is a supervised machine learning technique most commonly used in classification problems \citep{Cortes1995}. The algorithm is based on the concept of locating the hyperplane that best divides a labeled dataset into two classes\footnote{Note that the SVM algorithm is designed as a binary classifier, meaning that it can only distinguish between two classes. However, it is straightforward to extend the technique so that it can distinguish between multiple classes using the heuristic one-versus-rest (OVR) approach that splits the multi-class classification problem into multiple binary classification problems.} \citep{yu2012svm}. Fig.~\ref{Figure 3-1} shows a schematic illustration of a binary classification problem for the two classes marked with green circles and red diamonds. In this case, the data is perfectly separable. Note that an infinite number of hyperplanes can be defined that perfectly separate the two classes. The SVM algorithm now tries to find the one solution that maximizes the margin around the separating hyperplane. The width of the margin is thereby defined by the data points nearest to the separating hyperplane, which are called the support vectors. They define where the hyperplane is positioned, making them the most critical elements of the dataset.

Consider we have a labeled training set of $N$ samples with feature vectors $\mathbf{x}_i \in \mathbb{R}^{p} (i=1,N)$, where $p$ is the number of features, and corresponding labels $y_i \in \left\{-1, 1\right\} (i=1,N)$ specifying whether a given feature vector $\mathbf{x}_i$ belongs to the first class (i.e., $y_i=1$) or to the second class (i.e., $y_i=-1$). For the specific classification problem described in this paper, we have a total of 12 features (i.e., $p=12$) corresponding to the elements of the real and imaginary parts of the 6-C polarization vectors (Eqs~\ref{eq:1-2}--\ref{eq:1-6}). 

We now try to find a set of weights $\mathbf{w}\in \mathbb{R}^p$ and a constant factor $b$ that defines a hyperplane for which the prediction given by $\operatorname{sign}(\mathbf{w}^T\phi({\mathbf{x}_i})+b)$ is correct for most samples. The Support Vector Classification (SVC) algorithm now calls for the solution of the following optimisation problem \citep{hsu2003practical}: 

\begin{subequations}
\label{eq:svc}
\begin{alignat}{2}
&\! \min _{\mathbf{w}, b, \zeta} \frac{1}{2} \mathbf{w}^{T} \mathbf{w}+C \sum_{i=1}^{N} \zeta_{i}& \qquad &\\
&\text{ subject to } &{y}_{i}\left(\mathbf{w}^{T} \phi\left({\mathbf{x}_{i}}\right)+b\right) \geq 1-\zeta_{i},& \\
& &\zeta_{i} \geq 0, i=(1, \ldots, N).&
\end{alignat}
\end{subequations}

The minimization of $\frac{1}{2}\mathbf{w}^T\mathbf{w}$ will maximize the margin around the separating hyperplane. Since not all problems are perfectly separable (i.e., certain samples might fall on the wrong side of the plane), an additional penalty term is introduced when a sample is misclassified or falls within the margin. If all samples would be classified correctly, the term $y_i(\mathbf{w}^T\phi({\mathbf{x}_i})+b)$ would always be $\geq 1$. In reality, since the classes might not be perfectly separable, some samples are allowed to be at a distance $\zeta_i$ from their correct margin boundaries. The strength of the penalty is defined by the parameter $C$, which acts as an inverse regularization parameter. 

The function $\phi(\mathbf{x}_i)$ enables us to map the training features $\mathbf{x}_{i}$ into a higher dimensional space if the features are not linearly separable in the original feature space, thereby making the problem more flexible. In practice, it is not necessary to define the transformation function $\phi(\mathbf{x}_i)$ itself when implementing the optimization problem in Eq.~(\ref{eq:svc}), but it is sufficient to know the so-called kernel function $K(\mathbf{x}_i, \mathbf{x}_j)$ that computes the inner product of two feature vectors in the transformed space:

\begin{equation}
    K(\mathbf{x}_i, \mathbf{x}_j) = \phi(\mathbf{x}_i)^T\phi(\mathbf{x}_j)
\end{equation}

For particular choices of the function $\phi(\mathbf{x}_i)$, these inner products can be computed very efficiently. In this paper, we use the radial basis kernel function, which is defined with the free parameter $\gamma$ as:

\begin{equation}
K\left(\mathbf{x}_i, \mathbf{x}_j\right)=\exp \left(-\gamma\left\|\mathbf{x}_i-\mathbf{x}_j\right\|^{2}\right), \gamma>0
\end{equation}

\begin{figure}
    \centering
    \includegraphics[width=1.0\columnwidth]{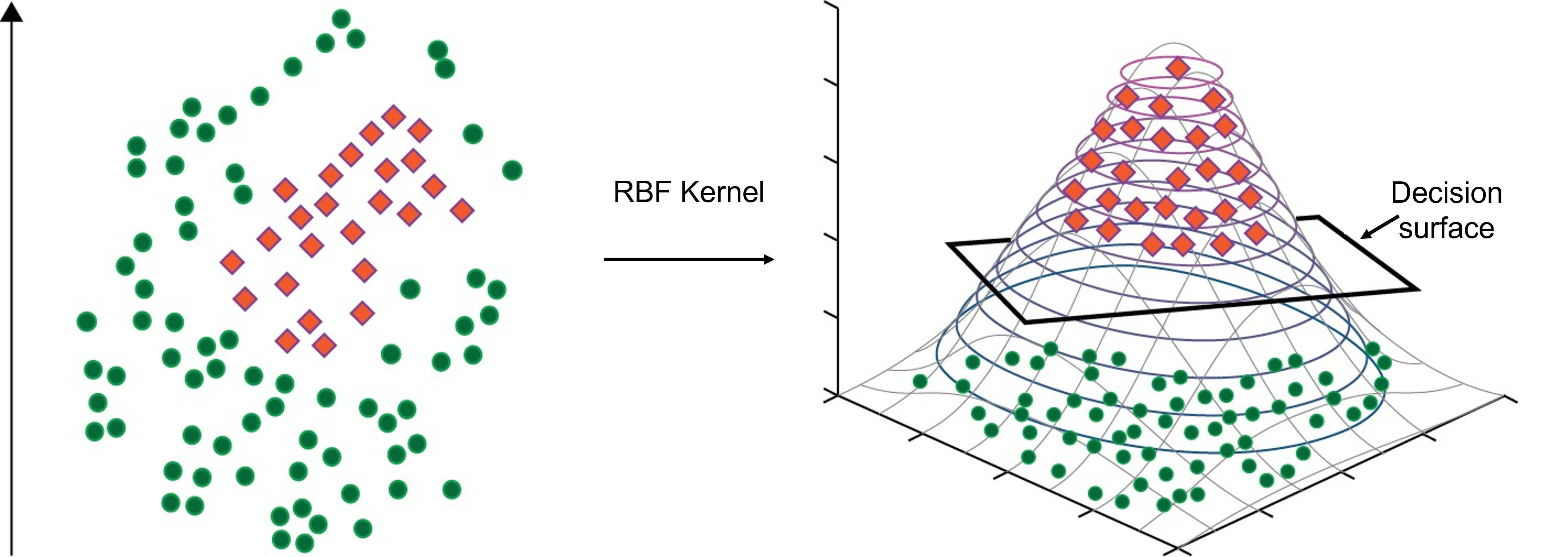}
    \caption{Schematic illustration of mapping the data from the original feature space where the data is not linearly separable, to a higher-dimensional space where the data is linearly separable.} 
    \label{Figure 3-2}
\end{figure}

Fig.~\ref{Figure 3-2} shows a dataset that is not linearly separable in the original feature space. After mapping of the data to a higher dimensional feature space using an appropriate mapping function $\phi(\mathbf{x}_i)$, the data becomes linearly separable.

\bsp 
     
\label{lastpage}

\end{document}